\documentclass{PoS}
\usepackage{epsfig,psfig}

\title{$\theta$ dependence of 4D SU($N$) gauge theories at finite
  temperature}

\ShortTitle{$\theta$ dependence of SU($N$) gauge theories}

\author{Claudio Bonati\\
  \\ Dipartimento di Fisica, Universit\'a di Pisa and INFN,
Largo Pontecorvo 2, I-56127 Pisa, Italy 
\\        E-mail: \email{bonati@df.unipi.it}}

\author{Massimo D'Elia\\
  \\ Dipartimento di Fisica, Universit\'a di Pisa and INFN,
Largo Pontecorvo 2, I-56127 Pisa, Italy \\
        E-mail: \email{delia@df.unipi.it}}

\author{Haralambos Panagopoulos\\
\\Department of Physics, University of Cyprus, Lefkosia,
CY-1678, Cyprus \\
        E-mail: \email{panagopoulos.haris@ucy.ac.cy}}

\author{\speaker{Ettore Vicari}
  \\ Dipartimento di Fisica, Universit\'a di Pisa and INFN,
Largo Pontecorvo 2, I-56127 Pisa, Italy \\
  E-mail: \email{vicari@df.unipi.it}}

\abstract{We report a study of the dependence of 4D SU($N$) gauge
  theories on the topological $\theta$ term at finite temperature, and
  in particular in the large-$N$ limit.  We show that the $\theta$
  dependence drastically changes across the deconfinement transition.
  The low-temperature phase is characterized by a large-$N$ scaling with
  $\theta/N$ as relevant variable, while in the high-temperature phase the
  free energy is essentially determined by the dilute instanton-gas
  approximation, with a simple $\theta$ dependence of the free-energy 
density  $F(\theta,T) - F(0,T) \sim 1 - \cos\theta$.   
}

\FullConference{
31st International Symposium on Lattice Field Theory - LATTICE 2013\\
July 29 - August 3, 2013\\	
		Mainz, Germany}

\begin{document}

4D SU($N$) gauge theories have a nontrivial dependence on the
topological $\theta$ term which can be added to the standard
Euclidean Lagrangian, i.e.
\begin{equation}
{\cal L}_\theta  = \frac{1}{4} F_{\mu\nu}^a(x)F_{\mu\nu}^a(x)
- i \theta q(x),\qquad
q(x)\equiv \frac{g^2}{64\pi^2} \epsilon_{\mu\nu\rho\sigma}
F_{\mu\nu}^a(x) F_{\rho\sigma}^a(x),
\label{lagrangian}
\end{equation}
where $q(x)$ is the topological charge density.  The $\theta$ term is
phenomenologically important, because it breaks both parity and time
reversal.  Its experimental upper bound within the strong-interaction
theory is very small, $|\theta| < 10^{-9}$~\cite{Baker}.
Nevertheless, the $\theta$ dependence is an interesting physical
issue, relevant to hadron phenomenology, an example being the
so-called U($1$)$_A$ problem.  Indeed, the nontrivial $\theta$
dependence provides an explanation to the fact that the U($1$)$_A$
symmetry of the classical theory is not realized in the hadron
spectrum~\cite{Hooft-74,Witten-79,Veneziano-79}.  The $\theta$
dependence at finite temperature ($T$) is related to the issue of the
effective restoration of the U($1$)$_A$ symmetry in strong
interactions at finite $T$, at high $T$ and around the chiral
transition, which may be also relevant to the nature of the transition
itself~\cite{PW-84,V-07}.

We report a study~\cite{BDPV-13} of the $\theta$ dependence of 4D
SU($N$) gauge theories at finite $T$, in particular across the
deconfining temperature $T_c$.  The finite-$T$ behavior is specified
by the free-energy density
\begin{equation}
F(\theta,T) = 
-\frac{1}{\cal V} \ln \int [dA] 
\exp \left(  - \int_0^{1/T} dt \int d^3 x\, {\cal L}_\theta \right),
\label{vftheta}
\end{equation}
where ${\cal V}=V/T$ is the Euclidean
space-time volume, and the gluon field satisfies $A_\mu(1/T,{\bf x}) =
A_\mu(0,{\bf x})$. The $\theta$ dependence can be parameterized as
\begin{eqnarray}
{\cal F}(\theta,T)\equiv F(\theta,T)-F(0,T)={1\over 2} \chi(T)
\theta^2 s(\theta,T),\label{ftheta}
\end{eqnarray}
where $\chi(T)$ is the topological susceptibility at $\theta=0$,
\begin{eqnarray}
&&\chi = \int d^4 x \langle q(x)q(0) \rangle_{\theta=0} 
= {\langle Q^2 \rangle_{\theta=0} \over {\cal V}},\label{chidef}
\end{eqnarray}
and $s(\theta,T)$ is a dimensionless even function of $\theta$ such
that $s(0,T)=1$.  Assuming analyticity at $\theta=0$, $s(\theta,T)$
can be expanded as
\begin{eqnarray}
s(\theta,T) = 1 + b_2(T) \theta^2 + b_4(T) \theta^4 + \cdots,
\label{stheta}
\end{eqnarray}
where only even powers of $\theta$ appear.

At $T=0$, where the free energy coincides with the ground-state
energy, large-$N$ scaling arguments~\cite{Hooft-74,Witten-98,VP-09}
applied to the Lagrangian (\ref{lagrangian}) indicate that the
relevant scaling variable is $\bar\theta\equiv {\theta/N}$, i.e.
${\cal F}(\theta) \approx N^2 {\cal G}(\bar{\theta})$ as $N\to
\infty$.  Comparing with Eq.~(\ref{ftheta}), this implies the
large-$N$ behavior
\begin{equation}
{\chi/\sigma^2}  = C_\infty + O(N^{-2}), \quad  
b_{2j}= \bar{b}_{2j}/N^{2j} + O(N^{-2j-2}),
\label{lnasyt0}
\end{equation}
where $\sigma$ is the string tension, $C_\infty$ and $\bar{b}_{2j}$
are large-$N$ constants. A nonzero value of $C_{\infty}$ is essential
to provide an explanation to the U($1$)$_A$ problem in the large-$N$
limit~\cite{Witten-79,Veneziano-79}.

The large-$N$ scaling (\ref{lnasyt0}) is not realized by the dilute
instanton gas (DIG) approximation. Indeed, at $T=0$, instanton
calculations fail due to the fact that large instantons are not
suppressed. On the other hand, the temperature acts as an infrared
regulator, so that the instanton-gas partition function is expected to
provide an effective approximation of finite-$T$ SU($N$) gauge
theories at high $T$~\cite{GPY-81}, high enough to make the overlap
between instantons negligible.  The corresponding $\theta$ dependence
is~\cite{GPY-81,CDG-78}
\begin{eqnarray}
&&{\cal F}(\theta,T) \equiv F(\theta,T)-F(0,T)
\approx \chi(T) \left( 1 - \cos\theta\right), 
\label{thdepht}\\
&&
\chi(T) \approx  T^4 \exp[-8\pi^2/g^2(T)] \sim  T^{-\frac{11}{3} N + 4}, 
\label{chitasy}
\end{eqnarray}
using $8 \pi^2/g^2(T) \approx (11/3) N \ln (T/\Lambda)+O(\ln\ln
T/\ln^2 T)$.  Therefore, the high-$T$ $\theta$ dependence
substantially differs from that at $T=0$\,: the relevant variable for
the instanton gas is just $\theta$, and not $\theta/N$.  The
DIG approximation also shows that $\chi(T)$, and therefore
the instanton density, gets exponentially suppressed in the large-$N$
regime, thus suggesting a rapid decrease of the topological activity
with increasing $N$ at high $T$. Since the instanton density gets
rapidly suppressed in the large-$N$ limit, making the probability of
instanton overlap negligible, the range of validity of the
DIG approximation is expected to rapidly extend toward
smaller and smaller temperatures with increasing $N$.  An interesting
question is how and when the DIG regime sets in.

In 4D SU($N$) gauge theories the low-$T$ and high-$T$ phases are
separated by a first-order deconfinement transition which becomes
stronger with increasing $N$ \cite{LTW-04}, with $T_c$ converging to a
finite large-$N$ limit:~\cite{LRR-12} $T_c/\sqrt{\sigma}= 0.545(2)+
O(N^{-2})$.  This suggests that the change from the low-$T$ large-$N$
scaling $\theta$ dependence to the high-$T$ DIG $\theta$ dependence
occurs around the deconfinement transition. See, e.g.,
Refs.~\cite{KPT-98,DPV-04,BL-07,PZ-08} for further discussions of this
scenario.

Due to the nonperturbative nature of the physics of $\theta$
dependence, quantitative assessments of this issue have largely
focused on the lattice formulation of the SU($N$) gauge theory, using
Monte Carlo (MC) simulations.  However, the complex character of the
$\theta$ term in the Euclidean QCD Lagrangian prohibits a direct MC
simulation at $\theta\ne 0$.  Information on the $\theta$ dependence
of physically relevant quantities, such as the ground state energy and
the spectrum, can be obtained by computing the coefficients of the
corresponding expansion around $\theta = 0$, which can be determined
by computing appropriate zero-momentum correlation functions of the
topological charge density
at $\theta=0$~\cite{DPV-02,DMPSV-06}. 
For example,
\begin{eqnarray}
\chi_l = {\langle Q^2 \rangle\over {\cal V}},\qquad
b_2 = -\, { \langle Q^4 \rangle - 3  \langle Q^2 \rangle^2  \over 
12 \langle Q^2 \rangle } , \qquad
b_4 =  {\langle Q^6 \rangle  -  15 \langle Q^2 \rangle \langle Q^4 \rangle  +
30 \langle Q^2 \rangle^3 
\over 360 \langle Q^2 \rangle} ,
\label{bbb}
\end{eqnarray}
where $Q$ is topological charge, 
$\chi_l$ is the the lattice topological susceptibility
($\chi_l\approx a^4 \chi$; $a$ is the lattice spacing).  The
coefficients $b_i$ in Eq.~(\ref{stheta}) are dimensionless and
renormalization-group invariant, therefore they approach their
continuum limit with $O(a^2)$ corrections.

We mention that issues related to $\theta$ dependence, particularly in
the large-$N$ limit, can also be addressed by other approaches, such
as AdS/CFT correspondence applied to nonsupersymmetric and
nonconformal theories, see
e.g. Refs.~\cite{Witten-98,PZ-08,AGMOO-00,kiritsis-etal}, and
semiclassical approximation of compactified gauge
theories~\cite{ZT-12, Unsal-12}.

\begin{table}
\begin{center}
\begin{tabular}{clll}
\hline
\multicolumn{1}{c}{$N$}&
\multicolumn{1}{c}{$\chi/\sigma^2$}&
\multicolumn{1}{c}{$b_2$}&
\multicolumn{1}{c}{$b_4$}\\
\hline
3 & 0.028(2) \cite{VP-09} 
& $-$0.026(3) \cite{PV-11}  &  0.000(1) \cite{PV-11} \\
4 & 0.0257(10) \cite{DPV-02} & $-$0.013(7) \cite{DPV-02} & \\
6 & 0.0236(10) \cite{DPV-02} & $-$0.008(4) \cite{BDPV-13} & 0.001(3) \cite{BDPV-13} \\
\hline
\end{tabular}
\caption{ Summary of known $T=0$ results for the ratio $\chi/\sigma^2$
  (where $\sigma$ is the $\theta=0$ string tension) and the first few
  coefficients $b_{2j}$ for $N=3,4,6$.  More complete reviews of
  results can be found in Refs.~\cite{VP-09,LP-12}; in particular
  other results for $b_2$ at $N=3$ are reported in
  Refs.\cite{DPV-02,Delia-03,GPT-07}. }
\label{t0res}
\end{center}
\end{table}

The large-$N$ scaling of the $\theta$ dependence is fully supported by
numerical computations exploiting the nonperturbative Wilson lattice
formulation of the 4D SU($N$) gauge theory at $T=0$, see, e.g., the
results reported in Table~\ref{t0res} for $N=3,4,6$ (see also
Refs.~\cite{VP-09,LP-12} for recent reviews).  A large-$N$
extrapolation of these data, using $a+b/N^2$ and $b/N^{2j}$ for
$\chi/\sigma^2$ and $b_{2j}$ respectively, leads to the estimates
\begin{equation}
C_\infty = \lim_{N\to\infty}
\chi/\sigma^2 = 0.022(2), \qquad  
\bar{b}_2 = \lim_{N\to\infty} N^2 b_2 = - 0.23(2).
\label{lnrs}
\end{equation}
This large-$N$ scenario is expected to remain stable against
sufficiently low temperatures.

\begin{figure}
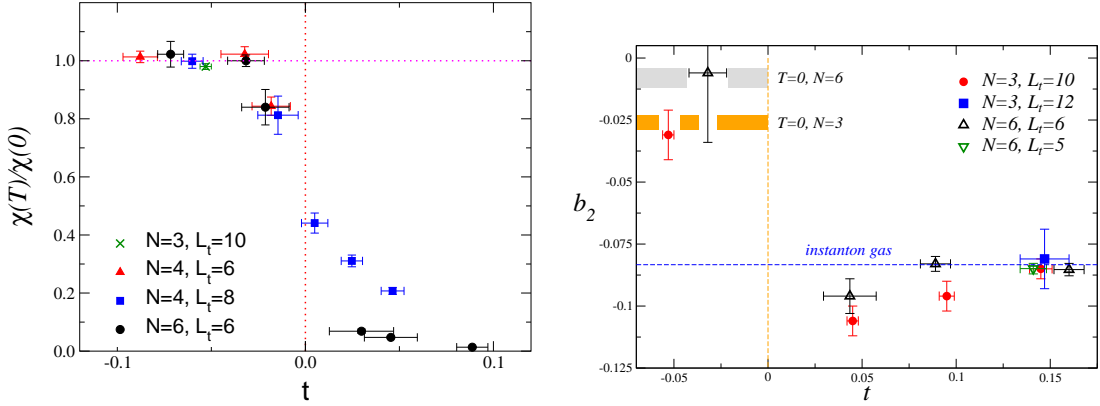

\begin{center}
\vskip 0.5truecm
{\begin{tabular}{cc}
\epsfig{width=7truecm,file=chit.eps} &
\hskip 0.1truecm
\epsfig{width=7truecm,file=b2t.eps}
\end{tabular}}
\caption{ The ratio $\chi(T)/\chi(0)$ between the topological
  susceptibility at $T$ and zero temperature (left) and the
  coefficient $b_2$ of the free-energy expansion around $\theta=0$
  (right), versus the reduced temperature $t\equiv T/T_c-1$, around
  the deconfinement transition corresponding to $t=0$.  We show data
  for various values of $N$ and lattice sizes $L_t\times L_s^3$ with
  $L_s/L_t\ge 4$, where $L_t,\,L_s$ are respectively the number of
  sites along the {\em time} and {\em space} directions.  The shadowed
  regions in the right figure indicate the $T=0$ estimates of $b_2$ for
  $N=3$ and $N=6$. The data for $N=4$ of the left figure are taken from 
Ref.~\cite{DPV-04}.
}
\label{tpt}
\end{center}
\end{figure}

The finite-$T$ lattice investigations of the large-$N$ behavior of
$\chi(T)$~\cite{LTW-05,DPV-04,GHS-02,ADG-97} indicate a nonvanishing
large-$N$ limit for $T<T_c$, remaining substantially unchanged in the
low-$T$ phase, from $T=0$ up to $T_c$.  Across $T_c$ a sharp change is
observed, and $\chi(T)$ appears largely suppressed in the high-$T$
phase $T>T_c$, in qualitative agreement with a high-$T$ scenario based
on the DIG approximation. Some MC data are shown in 
Fig.~\ref{tpt} (left panel).

A more stringent check of the actual scenario realized in 4D SU($N$)
gauge theories is provided by the higher-order terms of the expansion
(\ref{stheta}).  Indeed, the expansion coefficients $b_{2j}$ are
expected to scale like $N^{-2j}$ if the free energy is a function of
$\theta/N$ and to be $N$-independent in the DIG approximation, or,
more generally, if the relevant large-$N$ scaling variable is just
$\theta$.  In particular, the simple $\theta$ dependence of
Eq.~(\ref{thdepht}) may be observed at much smaller $T$ above $T_c$
with respect to the asymptotic one-loop behavior (\ref{chitasy}) of
$\chi(T)$ which is subject to logarithmic corrections.

We computed the first few coefficients of the expansion (\ref{stheta})
around $T_c$, for $N=3$ and $N=6$ to check the $N$ dependence, using
the lattice Wilson formulation of SU($N$) gauge theories, and a
smearing technique to determine the topological charge.  They require
high-statistics simulations due to the cancellation of volume factors
in their definitions (\ref{bbb}).  For details see
Ref.~\cite{BDPV-13}. Fig.~\ref{tpt} (right panel) shows the data for
$b_2$.  The MC results clearly show a change of regime in the $\theta$
dependence, from a low-$T$ phase where the susceptibility and the
coefficients of the $\theta$ expansion vary very little, to a high-$T$
phase where the coefficients $b_{2j}$ approach the instanton-gas
predictions.  In the high-$T$ phase they are definitely not consistent
with the large-$N$ scaling in Eq.~(\ref{lnasyt0}), which would imply a
factor of four in $b_2$, in going from $N=3$ to $N=6$.  On the other
hand, in the low-$T$ phase $b_2$ does not significantly differ from
the $T=0$ value.  This is consistent with the behaviour of the
topological susceptibility, see the left panel of Fig.~\ref{tpt}.
Although our MC results in the high-$T$ phase are obtained for
relatively small reduced temperatures $t\equiv T/T_c-1$, i.e. $t <
0.2$, the data for $b_2$ show a clear and rapid approach to the value
$b_2=-1/12$ of the instanton gas model for both $N=3$ and $N=6$, with
significant deviations visible only for $t\lesssim 0.1$.  The high-$T$
values of $b_2$ substantially differ from those of the low-$T$ phase,
and in particular from those at $T=0$ reported in Table~\ref{t0res}.
Also the estimates of $b_4$ are consistent with the small value
$b_4=1/360$.  The sharp behavior of the $\theta$ dependence at the
phase transition suggests that $T_c$ is actually a function of
$\theta/N$ at finite $\theta$, as put forward in Ref.~\cite{DN-12}.

A virial-like expansion can account for the deviations for $b_2$,
visible at $t \lesssim 0.1$, by correcting the asymptotic formula by a
term proportional to the square of the instanton density.  For
example, we may write
\begin{equation}
{\cal F}(\theta,T) \approx \chi (1-\cos\theta) + 
\chi^2 \kappa(\theta) + O(\chi^3),
\label{fthcorr}
\end{equation}
using the fact that $\chi(T)$ is proportional to the instanton
density, and $\kappa(\theta)$ can be parametrized as
$\kappa(\theta)=\sum_{k=2} c_{2k} \sin(\theta/2)^{2k}$.  The above
formula gives $b_2 \approx -{1/12} + {1\over 8}\, c_4 {\chi/T_c^4}$.
This predicts a rapid approach to the asymptotic value of the DIG
approximation, since $\chi$ gets rapidly suppressed in the high-$T$
phase, as suggested by Eq.~(\ref{chitasy}) and confirmed by the MC
results.  Moreover, a hard-core approximation of the instanton
interactions~\cite{CDG-78} gives rise to a negative correction, i.e.
$c_4<0$, explaining the approach from below to the perfect
instanton-gas value $b_2=-1/12$.

This numerical analysis provides strong evidence that the $\theta$
dependence of 4D SU($N$) gauge theory experiences a drastic change
across the deconfinement transition, from a low-$T$ phase
characterized by a large-$N$ scaling with $\theta/N$ as relevant
variable, to a high-$T$ phase where this scaling is lost and the free
energy is essentially determined by the DIG approximation,
which implies an analytic and periodic $\theta$ dependence. The
corresponding crossover around the transition becomes sharper with
increasing $N$, suggesting that the DIG regime sets in just above
$T_c$ at large $N$.

In full QCD the $\theta$ dependence is closely related to the
effective breaking of the U(1)$_A$ symmetry, through the axial anomaly
which is proportional to the topological charge density,
i.e. $\partial_\mu J_5^\mu(x) \propto {1\over N} q(x)$ in the chiral
limit. Its effects around the chiral transition may be relevant to the
nature of the transition itself. In the light-quark regime the nature
of the finite-temperature transition is essentially related to the
restoring of the chiral symmetry, and the corresponding symmetry
breaking pattern~\cite{PW-84}. In the relevant case of two light
flavors, this is ${\rm SU}(2)_L\otimes {\rm SU}(2)_R \rightarrow {\rm
  SU}(2)_V$, thus equivalent to O(4)$\rightarrow$O(3).  On the other
hand, if the effects of the axial anomaly are effectively suppressed
at the transition, the relevant symmetry breaking is ${\rm
  U}(2)_L\otimes {\rm U}(2)_R \rightarrow {\rm U}(2)_V $.  This
implies that, in the case of a continuous chiral transition (note
however that the transition may be also first order independently of
the symmetry breaking), the critical behavior belongs to different 3D
universality classes in the two cases~\cite{PV-13,BPV-05}.

Analogously to pure gauge theories, semiclassical instanton
calculations predict a substantial suppression of the instanton
density at large temperatures, $T\gg T_c$ say, where the DIG model is
expected to provide a reliable approximation~\cite{GPY-81}.  For
example, in QCD with two light flavors of mass $m$, the topological
susceptibility $\chi$ is expected to asymptotically decrease as $\chi
\sim m^2 \, T^{-\kappa}$, with $\kappa = {11\over 3}N - {16\over 3}$.
Although $\chi$ vanishes in the massless limit, the Dirac zero modes
associated with the instantons induce a residual contribution to the
${\rm U}(1)_A$ symmetry breaking, giving rise to a difference between
the susceptibilities of the so-called $\pi$ and $\delta$ channels at
high $T$,~\cite{Bazetal-12,CCCLMV-99} which behaves as
$\chi_\pi-\chi_\delta\sim T^{-\kappa}$ in the chiral limit.
Therefore, the DIG approximation suggests that the U(1)$_A$ symmetry
is not exactly recovered at finite $T$, although its breaking gets
largely suppressed with increasing the temperature.

The breaking of the ${\rm U}(1)_A$ symmetry at finite temperature has
been much investigated, even numerically by MC simulations of lattice
QCD, see e.g.
Refs.~\cite{Bazetal-12,CCCLMV-99,KLS-98,Bernard-etal-97,AFT-12,CAFHKMN-13,Buchoffetal-13}
and references therein.  These studies agree with a substantial
suppression of the ${\rm U}(1)_A$ anomaly effects at large
temperature, as predicted by the DIG model.  This scenario is
strenghtened by our numerical investigation of the pure ${\rm SU}(N)$
gauge theories.  However, the issue about the significance of this
suppression around the chiral transition is still debated.

\medskip

HP would like to thank the Research Promotion Foundation of Cyprus
for support, and INFN, Sezione di Pisa, for the kind hospitality.

\end{document}